\RequirePackage{lineno}
\documentclass[aps,prl,showpacs,twocolumn,preprintnumbers,amsmath,amssymb,,superscriptaddress,showkeys]{revtex4}

\usepackage{longtable}
\usepackage{graphicx}
\usepackage{dcolumn}
\usepackage{bm}
\usepackage{subfigure}

\usepackage[usenames]{color}
\begin{document}

\title{
On the nature of the spin polarization limit in the warped Dirac cone of the Bi$_2$Te$_3$
}
%
%
\author{A. Herdt}
\affiliation{%
Peter Gr\"{u}nberg Institut PGI-6, Forschungszentrum J\"{u}lich, D-52425 J\"{u}lich, Germany
}%
\affiliation{%
Fakult\"at f\"ur Physik and Center for Nanointegration Duisburg-Essen (CeNIDE), DE-47048 Duisburg, Germany
}%

\author{L. Plucinski}
\email{l.plucinski@fz-juelich.de}
\affiliation{%
Peter Gr\"{u}nberg Institut PGI-6, Forschungszentrum J\"{u}lich, D-52425 J\"{u}lich, Germany
}%
\affiliation{%
Fakult\"at f\"ur Physik and Center for Nanointegration Duisburg-Essen (CeNIDE), DE-47048 Duisburg, Germany
}%
\affiliation{%
J\"{u}lich Aachen Research Alliance - Fundamentals of Future Information Technologies (JARA-FIT), Germany
}%

\author{G. Bihlmayer}
\affiliation{%
Peter Gr\"{u}nberg Institut PGI-1 and Institute for Advanced Simulation, Forschungszentrum J\"{u}lich, D-52425 J\"{u}lich, Germany
}%

\author{G. Mussler}
\affiliation{%
J\"{u}lich Aachen Research Alliance - Fundamentals of Future Information Technologies (JARA-FIT), Germany
}%

\affiliation{%
Peter Gr\"{u}nberg Institut PGI-9, Forschungszentrum J\"{u}lich, D-52425 J\"{u}lich, Germany
}%

\author{S. D\"{o}ring}
\affiliation{%
Fakult\"at f\"ur Physik and Center for Nanointegration Duisburg-Essen (CeNIDE), DE-47048 Duisburg, Germany
}%

\author{J. Krumrain}
\affiliation{%
J\"{u}lich Aachen Research Alliance - Fundamentals of Future Information Technologies (JARA-FIT), Germany
}%
\affiliation{%
Peter Gr\"{u}nberg Institut PGI-9, Forschungszentrum J\"{u}lich, D-52425 J\"{u}lich,
Germany
}%

\author{D. Gr\"{u}tzmacher}
\affiliation{%
Peter Gr\"{u}nberg Institut PGI-9, Forschungszentrum J\"{u}lich, D-52425 J\"{u}lich,
Germany
}%
\affiliation{%
J\"{u}lich Aachen Research Alliance - Fundamentals of Future Information Technologies (JARA-FIT), Germany
}%

\author{S. Bl\"{u}gel}
\affiliation{%
Peter Gr\"{u}nberg Institut PGI-1 and Institute for Advanced Simulation, Forschungszentrum J\"{u}lich, D-52425 J\"{u}lich, Germany
}%

\author{C. M. Schneider}
\affiliation{%
Peter Gr\"{u}nberg Institut PGI-6, Forschungszentrum J\"{u}lich, D-52425 J\"{u}lich, Germany
}
\affiliation{%
Fakult\"at f\"ur Physik and Center for Nanointegration Duisburg-Essen (CeNIDE), DE-47048 Duisburg, Germany
}%
\affiliation{%
J\"{u}lich Aachen Research Alliance - Fundamentals of Future Information Technologies (JARA-FIT), Germany
}%

\date{\today}

\begin{abstract}
The magnitude of electron spin polarization in topologically protected surface states is an important parameter with respect to spintronics applications. In order to analyze the warped spin texture in Bi$_2$Te$_3$ thin films, we combine angle- and spin-resolved photoemission experiments with theoretical \textit{ab initio} calculations. We find an \textit{in-plane} spin polarization of up to $\sim$~45\% in the topologically protected Dirac cone states near the Fermi level. The Fermi surface of the Dirac cone state is warped and shows an \textit{out-of-plane} spin polarization of $\sim$~15\%. These findings are in quantitative agreement with dedicated simulations which find electron density of the Dirac cone delocalized over the first quintuple layer with spin reversal occurring in the surface atomic layer.
\end{abstract}

\pacs{000}
\maketitle

Gapless surface states in the family of three-dimensional (3D) topological insulators (TIs) Bi$_2$Te$_3$, Bi$_2$Se$_3$, Sb$_2$Te$_3$, and their alloys have recently attracted considerable attention due to their potential in producing fully spin-polarized currents for spin-electronic applications \cite{Zhang2009NP}. Theoretical models predict that these topological surface states (TSS) are \emph{topologically protected} and fully spin-polarized. A simple explanation of the Bi$_2$Te$_3$ surface electronic structure properties comes from symmetry arguments and the topological theory. In the bulk band structure near the $\Gamma$ point, the band character order is inverted due to the spin-orbit interaction. At the boundary of the crystal, this leads to the formation of gapless edge-states that are protected by time-reversal symmetry. This can be classified in terms of the topological Z$_2$ invariant \cite{Kane2005}, which is used to classify the Quantum Spin Hall effect, and it results in the prediction of fully spin-polarized single-branched Dirac cones \cite{Kane2005a} as the surface states of the 3D TIs. A more detailed analysis of such states can be performed by using density functional theory (DFT). Within this approach, the average spin polarization of the TSS was predicted to be strongly reduced due to spin-orbit entanglement \cite{Yazyev2010PRL}, and the vector of spin polarization was found to change its orientation between the subsequent atomic layers \cite{Eremeev2012NatComm,Henk2012PRL}.

The combination of angle- and spin-resolved photoemission spectroscopy (SP-ARPES) with \emph{ab-initio} theoretical calculations is an efficient approach to investigate the exotic spin texture of the TSS \cite{Pan2011PRL,Jozwiak2011PRB}. Recently, experimental investigations on the warped spin texture of Bi$_2$Te$_3$ were reported from bulk samples, finding the experimental spin polarization in the ensemble of the photoelectrons ranging from 20\% (Hsieh \emph{et al.} \cite{Hsieh2009Nature}) to 60\% (Souma \emph{et al.} \cite{Souma2011PRL}), however, they were to date not confirmed in thin films.

Since the implementation into spintronic devices asks for thin film structures, for which the promising properties observed on bulk single crystals have not yet been demonstrated, a thorough understanding of the spin polarization of the electronic states in thin films is mandatory for a successful engineering of devices. In order to establish a relation to DFT calculations, the experimental studies must be based on high-quality TI thin films.

In this work, we investigate the spin polarization behaviour of epitaxial thin films of the narrow gap semiconductor Bi$_2$Te$_3$ by SP-ARPES. We determined the spin texture within the warped Fermi surface finding a maximum value for the spin polarization vector of $\sim 45$\% in the \textit{in-plane} and $\sim 15$\% in the \textit{out-of-plane} component in the Dirac cone photoelectrons. The experimental results agree well with the spectral weights derived from our calculations.

The spin-polarized photoelectron spectroscopy experiment has been carried out at Beamline~5 of the 1.5~GeV synchrotron radiation source DELTA (Dortmund, Germany) using linearly polarized light (photon energy $h\nu = 24$~eV) at an overall resolution of \mbox{150~meV \cite{Plucinski2010instr}.} The experimental end-station includes a commercial Scienta SES-2002 hemispherical spectrometer equipped with a combination of an optimized high transmission spin-polarized low-energy electron diffraction (SPLEED) based detector \cite{Kirschner1979PRL,Yu2007SS} and a two-dimensional delay-line-detector (DLD) system \cite{Plucinski2010instr}, and allows
to simultaneously measure one of the \emph{in-plane} and the \emph{out-of-plane} components of the spin polarization vector $\textit{\textbf{P}}$. We have performed angle- and spin-polarized PES measurements on epitaxial MBE-grown thin film Bi$_2$Te$_3$ epilayers deposited on \textit{n}-type Si(111) substrates, with detailed growth and film characterization described in Ref. \onlinecite{Krumrain2011JoCG}. After being exposed to air, the samples were cleaned by Ar-ion sputtering and annealing cycles under ultra high vacuum. A complete surface cleaning procedure is presented in Ref. \onlinecite{Plucinski2011APL}. The ARPES spectra were measured on samples kept at 15~K with overall resolution of 20~meV at the laboratory based system using a He I discharge source ($h\nu=21.22$~eV) and a Scienta SES-200 spectrometer \cite{Suga2010RSI}. The thickness of the samples after the cleaning procedure was between 10 nm and 40 nm and we have obtained consistent results on several samples.

\begin{figure}[h!]
\includegraphics[width=9cm]{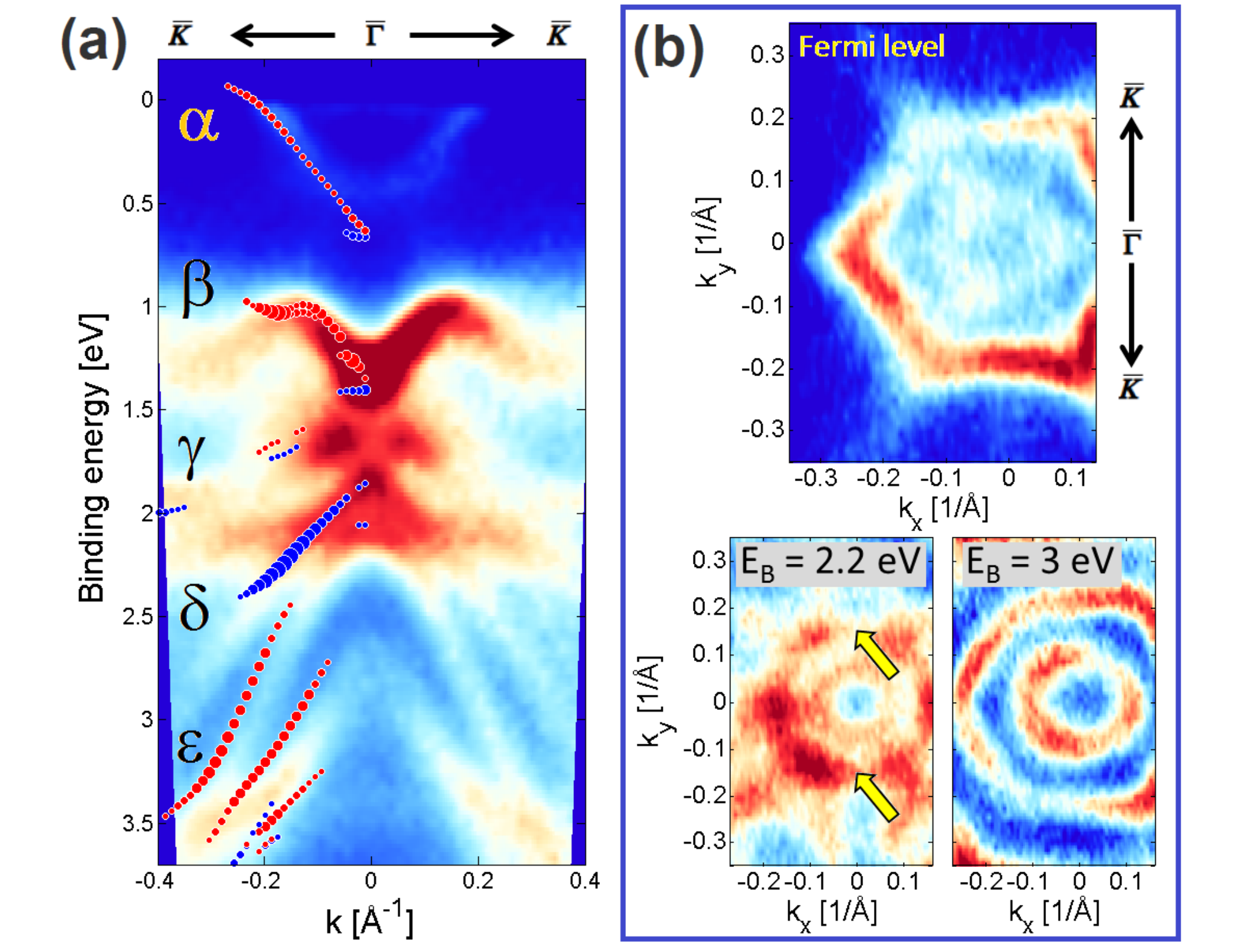}
\caption{\label{fig:ARPES} (Color online) (a) Photoemission map of the 40 nm Bi$_2$Te$_3$ film taken at 15~K using the He I (h$\nu \approx 21.22$~eV) excitation along the $\overline{\Gamma K}$ direction, with the overlaid DFT \textit{in-plane} calculation (see Fig.~\ref{fig:theo1}~(a)), shifted by 0.45~eV to achieve the best match to the experiment. Blue and red solid circles indicate the spin direction, whereas the size of the symbols is related to the weight of the states at the surface. (b) Constant energy surfaces for the Fermi level (top), $E_B = 2.2$~eV and $E_B = 3.0$~eV. Each map was separately normalized for optimized contrast. Arrows in the $E_B = 2.2$~eV map indicate the two example symmetric positions on the ring, related to the $\delta$ band (see panel (a)), for which the direction of the polarization vector is reversed due to its helical nature.}
\end{figure}

Fig.~\ref{fig:ARPES} shows high resolution ARPES results from a Bi$_2$Te$_3$ film measured under these conditions. The valence band structure in panel (a) depicts that the spectral weight is dominated by the set of strongly dispersing bands between 1 and 3.5~eV binding energy $E_B$. Panel (b) shows three distinct constant energy maps, where well defined features indicate a high crystalline quality of the prepared surfaces.

Clear Dirac cones are present in the ARPES maps. However, they do not show a distinct Dirac point at $\overline\Gamma$, as reported in our previous work \cite{Plucinski2011APL}. Compared to cleaved single crystals \cite{Chen2009} our cones are wider, with smaller Fermi velocity, similar to the states measured on air-exposed surfaces \cite{Chen2012PNAS}, which can be explained by cleaning-induced surface disorder and the intercalation of foreign atoms into the so-called van der Waals gap between the QLs \cite{Ye2011a}. The warped Fermi surface hexagram Dirac cone structure of Bi$_2$Te$_3$ \cite{Chen2009,Fu2009PRL}, which exhibits a sizeable \textit{out-of-plane} component, is a hallmark of this material. Moreover, an additional contribution from the bulk conduction band with a trigonal symmetry is present. Clear trigonal symmetry is also present in the map taken at $E_B = 3$~eV, indicating that the $60^\circ$ rotated crystalline domains \cite{Li2010NJP} are not present in our surface. Therefore, the \textit{out-of-plane} spin polarization component of the warped Dirac cone \cite{Souma2011PRL,Xu2011a}, which inverts its sign every $60^\circ$, could be observed in the spin-polarized photoemission measurement as will be shown next.

\begin{figure*}[ht!]
\includegraphics[width=18cm]{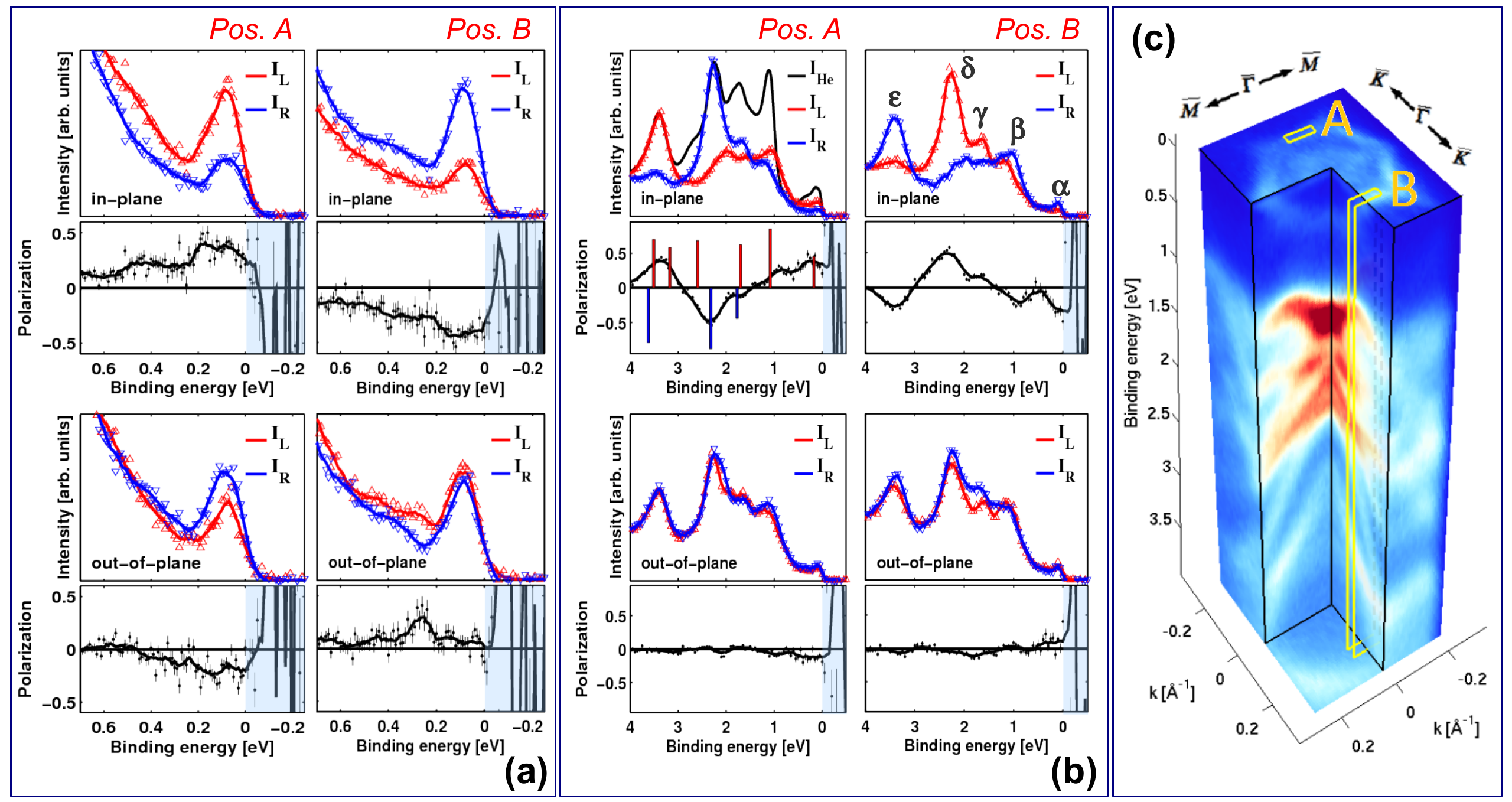}
\caption{\label{fig:SPLEED_NE} (Color online) Spin-polarized data taken (a) near the Fermi level and (b) at higher binding energies on selected \textit{\textbf{k}}-space locations along the $\overline {\Gamma  K}$ direction on the 40 nm Bi$_2$Te$_3$ film. Here, blue and red solid lines show smoothed $I_L$ and $I_R$ intensities, respectively.  States being described within the text are indicated by letters $\alpha$ to $\epsilon$. The top row indicates the \textit{in-plane} spin-vector component intensities, whereas the \textit{out-of-plane} intensities are plotted in the lower rows. The deduced spin-asymmetries $P_x$ and $P_z$ are plotted below the corresponding intensity plots with standard deviations given by vertical error bars, whereas the solid line represents smoothed data. Furthermore, panel (b) shows the \emph{He I} data integrated over the SPLEED entrance (the black curve) and the theoretical prediction for the \textit{in-plane} spin polarization in surface-related eigenvalues at $|k_{\parallel}|=0.18$~\AA$^{-1}$ shifted by 0.45~eV to achieve the best match to the experiment (the bar graph). (c) Experimental three-dimensional illustration of the band structure of a Bi$_2$Te$_3$ thin film over the full valence band region, indicating the $\textit{\textbf{k}}$-space volumes A and B which are integrated in the spin-polarized experiment.}
\end{figure*}

Spin-polarized photoemission measurements performed at 200~K are presented in Figs.~\ref{fig:SPLEED_NE}~(a) and (b), while Fig.~\ref{fig:SPLEED_NE}~(c) illustrates the $\textit{\textbf{k}}$-space volume over which the spin-polarized spectra are integrated \cite{Plucinski2010instr}. The polarization $P=A/S$ with the asymmetry $A=(I_L-I_R)/(I_L+I_R)$, in which $I_L$ and $I_R$ are the signals for the beams scattered from the W(001) crystal in opposite mirror directions, was computed using the Sherman function $S=0.25$ \cite{Kirschner1979PRL,Yu2007SS}. In analogy to the analysis shown in \cite{Jozwiak2011PRB}, the unpolarized constant background above $E_F$ has been removed. The combination of our cleaning procedure and the choice of the photon energy minimized the spectral weight related to the conduction band states, allowing our spin-polarized spectra to be free from the un-polarized background in the Dirac cone region.

For the \emph{in-plane} spectra we have obtained up to \mbox{$P_x \sim 45$\%} in the Dirac cone and in the wider range up to \mbox{$\sim 55$\%} for the most pronounced $\delta$ feature at $E_B \simeq 2.3$~eV. The \textit{out-of-plane} spin component is much smaller, and the confirmation is based on the standard deviation analysis and the polarization reversal between points (A) and (B). Based on this we obtained $P_z \sim 15$\% polarization in the Dirac cone $\alpha$ state. There are also indications of sizeable \textit{out-of-plane} polarization in other states, especially for the $\gamma$-state which has the opposite polarity to the Dirac cone, however, in the current data this is close to the noise limit.

The $\textbf{\textit{P}}$ value extracted from the experimental data can be affected by the instrumental asymmetry of the spin polarimeter. In typical spin-polarized photoemission experiments on ferromagnetic thin films this issue is addressed by re-magnetizing the film in the opposite direction to effectively cancel instrument related asymmetries. However, in a non-ferromagnetic material such as Bi$_2$Te$_3$ magnetization is not possible, and one has to rely on the absolute calibration of the spin polarimeter. Instead, the spectra have been measured on the two precisely opposite sides of the warped Fermi surface Dirac cone rim, which shows the full reversal of spin polarization vector due to the helical nature of the Dirac cone. Our calibration is confirmed by comparing the spectra measured in positions (A) and (B) as shown in Fig.~\ref{fig:SPLEED_NE}, where clear reversal is observed in all cases, with virtually the same results on the two \textit{\textbf{k}}-space points.

\begin{figure*}[!ht]
\includegraphics[width=18cm]{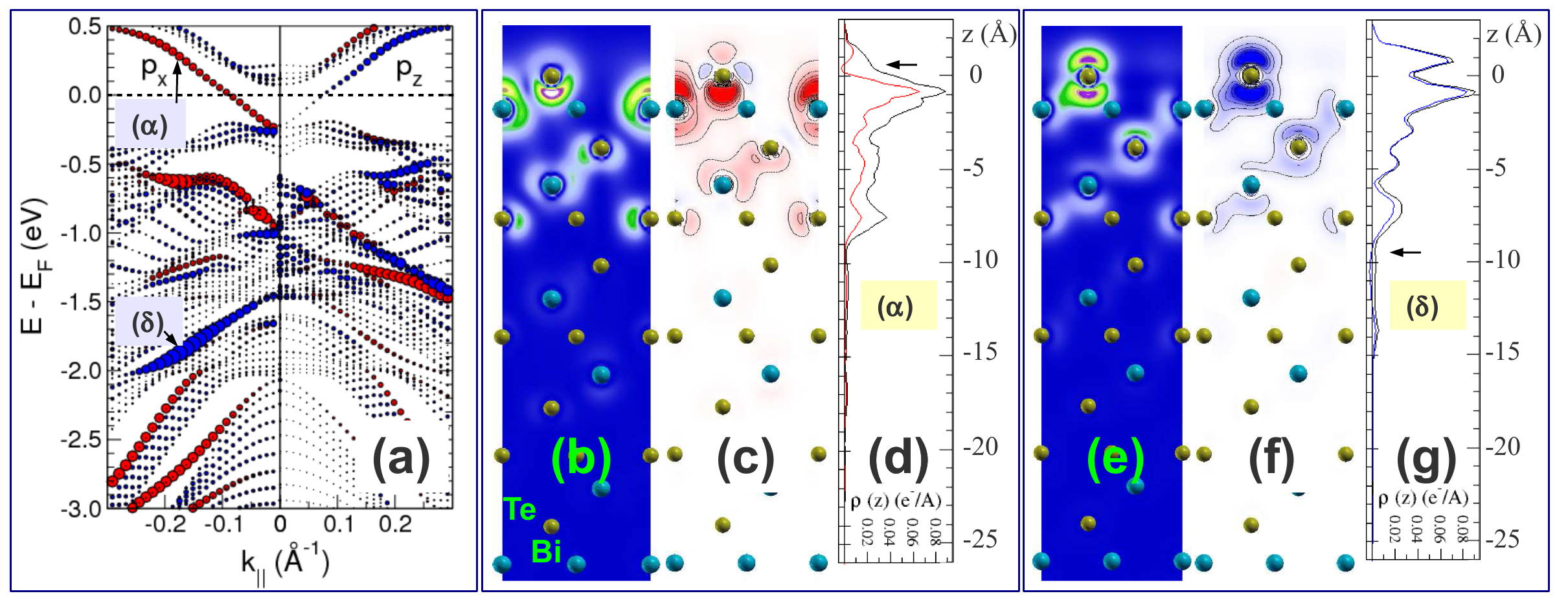}
\caption{\label{fig:theo1} (Color online) (a) Simulated spin-polarized spectral weight of the 6~QL Bi$_2$Te$_3$ slab along the $\overline{\Gamma K}$ direction. The size of the circles gives the absolute spin polarization of the states for the \emph{in-plane} $P_x$ and \emph{out-of-plane} $P_z$ components. Herein, the photoelectron mean free path has been taken into account in reference to the experiment. Two states, namely $\alpha$ and $\delta$ analyzed in the further drawings are indicated by arrows. (b) and (e) show charge, and (c) and (f) spin densities plotted in real space integrated over the $(x,y)$ plane for the $\alpha$ and $\delta$ state at $|k_{\parallel}|=0.18$~\AA$^{-1}$, respectively. (d) and (g) show charge and spin densities for those states with respect to the distance from the surface. Arrows in (d,g) denote spin reversals taking place for each state. See the text for details. The Dirac-cone state has a smaller polarization, due to the polarization inversion above the topmost Te atom (local spin-reversal). The \textit{in-plane} polarization value integrated taking into account the photoemission information depth is about 45\%, in contrast to the surface state $\delta$ that has 88\%.}
\end{figure*}

In order to perform a consistent analysis of the photoemission experiment and comparison to DFT calculations, the electron scattering length has to be taken into account. In the kinetic energy range related to the VUV photon energy range, which was so far used in all the spin-polarized photoemission experiments on Bi$_2$Te$_3$ and similar materials, the mean free path of the electrons is in the order of 10~\AA~(a single QL), or less. Nevertheless, the resulting measured polarization $\textbf{\textit{P}}$ is a function of the symmetry of the particular states, the light polarization, and the incidence angle of the beam \cite{Park2012PRL}. We have performed a dedicated simulation of the electronic properties of the Bi$_2$Te$_3$ thin film system using the full-potential linearized augmented plane wave (FLAPW) method implemented in the FLEUR code (for details see Ref.~\onlinecite{flapw}). Several distinct spin-polarized states, localized in the surface QL, are present within the valence band as shown in Fig.~\ref{fig:theo1}~(a). The calculated local depth-resolved spin contributions are given in Figs.~\ref{fig:theo1}~(b)-(g) for $|k_{\parallel}|=0.18$~\AA$^{-1}$ corresponding to the center position of the spin detector integration area as shown in Fig.~\ref{fig:SPLEED_NE}~(c). In particular, we have calculated the spin density for the atom $\eta$ as $\rho^{\eta}_{k_{\parallel}}=(n^{\uparrow}_{\eta}-n^{\downarrow}_{\eta})$ and the corresponding initial spin polarization values including the photoelectron mean free path $\lambda$ \cite{note} using $P=\frac{\sum_{\eta}{e^ {-z_{\eta}\lambda}(n^{\uparrow}_{\eta}-n^{\downarrow}_{\eta})}}{\sum_{\eta}{e^ {-z_{\eta}\lambda}(n^{\uparrow}_{\eta}+n^{\downarrow}_{\eta})}}$. The direction and magnitude of the spin polarization vectors of the corresponding states depend on the distance from the surface, with clear spin reversal taking place within the 2~\AA~surface region in case of the Dirac cone state $\alpha$, as shown in Fig.~\ref{fig:theo1}~(d), in agreement to a previous study \cite{Eremeev2012NatComm}. The spin polarization integration over the surface QL is the highest for the surface state indicated by $\delta$ in Fig.~\ref{fig:theo1}~(a). Its spin density with respect to the distance from the surface is plotted in Fig.~\ref{fig:theo1}~(g). Unlike the Dirac cone, in this state only a tiny spin reversal takes place in the bottom layer of the surface QL, $\sim 9$~\AA~below the surface. The difference between the Dirac cone state and the aforementioned $\delta$ state is further depicted in charge and spin densities plotted in Figs.~\ref{fig:theo1}~(b,c) and (e,f), where clear spin reversal at the surface atom is observed for the Dirac cone state, while not observed for the $\delta$ state. Furthermore, one can clearly observe the strong, surface Te atom-localized, $p_z$ character of this surface state, which has a Rashba-type spin polarization. In contrast, the Dirac cone state has a hybridized nature, with most charge related to the $p_z$ state of the second layer Bi atom, but with a significant contribution of the reversed spin from the surface layer Te atom. The helicity of the surface features in Bi$_2$Te$_3$ is not only a feature of the Dirac cone topological state, it is a general feature for all surface-related states, while another feature is the existence of a significant \emph{out-of-plane} spin polarization vector component, predicted for some of these states. Our experimental \textit{in-plane} results from Fig. \ref{fig:SPLEED_NE} present a stunning agreement with the theoretical prediction from Figs.~\ref{fig:ARPES}~(a) and \ref{fig:theo1}~(a), with polarization direction of the bands following the predicted order, i.e. $\alpha$, $\beta$ and $\epsilon$ are polarized opposite to $\gamma$ and $\delta$. The relative magnitude of the polarization of these features is also in unison with the theory, as is the helical nature of all the polarized states, confirmed by the spin reversal measured at surface Brillouin zone locations (A) and (B) shown in Fig. \ref{fig:SPLEED_NE}. The predicted 88\% polarization of the $\delta$ state cannot be directly observed in experiment since at higher $E_B$ spectral features are broadened and even in the presence of local gaps, the spectral weight contains significant contributions from the bulk band structure reducing the experimentally observable polarization to $\approx$ 55\%.

In conclusion we have confirmed that the topological protection mechanism in Bi$_2$Te$_3$ thin film samples prepared by an optimized \textit{in-situ} procedure leads to helical single branched warped Dirac cone states. Our experiments show the strong Fermi surface warping predicted by theory, with both \textit{in-plane} and \textit{out-of-plane} spin polarization components observed in the spin-polarized data, satisfying the antisymmetric spin property \mbox{$\boldsymbol \sigma (\textit{\textbf{k}})=-\boldsymbol \sigma(-\textit{\textbf{k}})$.} The Dirac cone state is delocalized over the surface QL, and its spin orientation changes within subsequent layers, including the \textit{in-plane} spin reversal over the first atomic layer. Despite the complications that might arise in the interpretation of photoemission spectra \cite{Park2012PRL} we notice that the experimental $P_x$ value is in a good quantitative agreement with the theoretical prediction of 45\%, when  including the photoelectron information depth.

Applications in spintronics require high spin polarization in the Dirac state and one of the possible ways to increase the overall spin polarization in the Bi$_2$Te$_3$ Dirac cone is to induce higher surface localization of the Dirac cone states, possibly ceasing any significant spin polarization inversions. This can be achieved by the deposition of thin layers of selected elements, and it was recently shown that a Bi-bilayer introduces higher localization of the Dirac cone state \cite{Hirahara2011PRL}, which suggests that the spin polarization in such states might be increased.

We acknowledge stimulating discussions with Caitlin Morgan on the course of editing the manuscript. This work is supported by a grant from the NRW Research School "Research with Synchrotron Radiation" funded by the Northrhine-Westphalia Ministry for Innovation, Science, Research, and Technology (Grant No. 321.2-8.03.06-58782).


\begin{thebibliography}{26}
\bibitem[{not()}]{note}
\bibinfo{note}{We have used an electron mean free path of $\lambda=10$ \AA.}
\end{thebibliography}

\end{document}